# Intelligent Sensor based Bayesian Neural Network for Combined Parameters and States Estimation of a Brushed DC Motor

Hacene MELLAH
Electrical Engineering Department,
Ferhat Abbas Sétif 1 University,
LAS laboratory, Sétif, Algeria

Kamel Eddine HEMSAS
Electrical Engineering Department,
Ferhat Abbas Sétif 1 University,
LAS laboratory, Sétif, Algeria

Rachid TALEB
Electrical Engineering Department,
Benbouali Hassiba University of
Chlef, Chlef, Algeria

***Abstract*—The objective of this paper is to develop an Artificial Neural Network (ANN) model to estimate simultaneously, parameters and state of a brushed DC machine. The proposed ANN estimator is novel in the sense that his estimates simultaneously temperature, speed and rotor resistance based only on the measurement of the voltage and current inputs. Many types of ANN estimators have been designed by a lot of researchers during the last two decades. Each type is designed for a specific application. The thermal behavior of the motor is very slow, which leads to large amounts of data sets. The standard ANN use often Multi-Layer Perceptron (MLP) with Levenberg-Marquardt Backpropagation (LMBP), among the limits of LMBP in the case of large number of data, so the use of MLP based on LMBP is no longer valid in our case. As solution, we propose the use of Cascade-Forward Neural Network (CFNN) based Bayesian Regulation backpropagation (BRBP). To test our estimator robustness a random white-Gaussian noise has been added to the sets. The proposed estimator is in our viewpoint accurate and robust.**

## I. INTRODUCTION

We said that when we can measure a physical quantity, we know something about it, but when we cannot quantify it, our knowledge about it is very poor and insufficient, so without quantifying science does not advance.

The DC motor speed controllers frequently use feedback from a speed measuring device, such as a tachometer or an optical encoder [1,2], but this later, adds an additional cost and congestion throughout the installation [2,3], the problems related to the speed measurement are detailed in the [3].

The simplest estimation method is based on the steady-state voltage equation, where the speed is written as a function of armature voltage and current; the peaks due to converter especially in the transient state affect this speed and the link resistance-temperature is ignored on the other hand, it is the two major inconvenient of this method [1].

R. Welch Jr. *et all* [4] discuss the temperature effects on electrical and mechanical time constants, he prove that these time constants are not constant value, in addition the motor's electrical resistance and its back EMF are depend on temperature.

In [5-8] we find several methods about DC machine temperature measurement, but the problems of temperature measurement are more complicated and difficult to solve than the speed measurement problems, since, the rotor is in rotation. The temperature variation is strongly nonlinear depend on the load, the supply quality, the cooling conditions, the design and the environment conditions. Actually, the problems of armature temperature measurement are not totally resolved.

In literature [9-10], a finite element method (FEM) was usually used to obtain generally a 3D thermal distribution in all electrical machine point. The major advantage of this method is that is suitable to help a designer to optimize the cost, weight and cooling mode in the goal to increase the efficiency and motor's lifetime [10], generally, the FEM is hard to implement in real time both for the control or monitoring, on the other hand, this approach has an enormous resolution time.

According the literature [11-15], we can distinguish two types of electrical machines thermal modeling approaches:

The first one is thermal model-based approaches, this approach based to divide the machine into homogeneous components unscrewed in order to ensure each part has uniform thermal characteristics such as thermal capacitors, thermal resistances and heat transfer coefficients [11, 15]. The identification of the model is performed either by the finite element technique or by a high range of temperature measurement. These models are generally very detailed so, too complex for real time application [16], however, many researchers simplify this model for the real time applications [15, 17]. This approach is robust, unfortunately this model is not generalized and a few measurements are needed for each motor [11, 16].

The second one is the parameter-based approaches, this approach based to get the temperature from the online resistance estimation [12-14] or identified [18-19]. Therefore, the estimate temperature takes under consideration the thermal environmental conditions. This method can respond to changes in the cooling conditions, and is accurate, but it is generally too sensitive [20].

This work was supported in part by Electrical Engineering Department, Ferhat Abbas Setif1 University (UFAS1) and in other part by Algerian ministry of research and High education.

P.P. Acarnley *et all* [1] proposes an Extended Kalman Filter (EKF) is implemented to estimate both the speed and armature temperature, but the EKF has problems with the matrices initialization step for each machine, therefore, the risk of divergence is not very far and not forgotten its dependence of the mathematical model.

R. Pantonial *et all* [21] propose the using of EKF in two steps, the first one is in the steady-state used to estimate the electromechanical behavior, and the second one is a transient version used to estimate the thermal behavior. However, in this case, the system is decoupled and the temperature effect on the resistance is not into account for the steady-state model.

A new nonlinear estimation strategy is proposed in the recent paper in this field based on combining elements of the EKF with the smooth variable structure filter (SVSF) to estimate the stator winding resistance [22], in this research we find only a resistance estimation approach, also the link temperature-resistance is ignored, then this is the simplest estimator version.

M. Jabri *et all* use a fuzzy logic technic to estimate the field and armature resistance of DC series motor, this is an important problem in order to implement a robust closed loop control [23], in their newest version [24], present a comparative study between a Levenberg-Marquardt (LM) and LM with tuning Genetic Algorithms (GA) to adjust relaxation. However, in the two versions, only the resistance and the flux were estimated and the link temperature-resistance is ignored.

The most important electrical machine parameter is the winding temperature, the winding temperature affects both the machine's lifetime and accuracy of control, when the winding temperature is equal or superior to the supported winding insulation temperature, this critical temperature affect directly on the machine lifetime; thus, good knowledge of the thermal state of the machine is very important.

In this context, obtaining the temperature by brittle, expensive sensors and adds a congestion to the overall installation, without forgetting the problematic of the sensor placement, therefore, the sensor is not the right solution [16]. In addition, using a Kalman filter, which is difficult to stabilize and the problematic of covariance matrices choices, remains the two major inconveniences, we propose an intelligent universal estimator based on ANN.

The ANN widely used in different engineering domain, such as renewable energy [25], chemical [26], pharmaceutical [27 ] and mechanics[28], as well the ANN used in several engineering applications such as control [29], optimization [30], modeling [31] and condition monitoring [32]. In addition, the ANN can used alone [33] or mixed with other technic such as GA [34], Particle Swarm Optimization (PSO) [35] and Fuzzy Logic [36].

One of the most commonly phrased questions in neural computation techniques refers to the size of the network that provides the best results. Although various ‘‘hints and tips’’ like suggestions have been pointed out so far, there is still no clear answer to reply to this question [37,38]. This paper describes and applies an intelligent technique for combined speed, temperature and resistance estimation in a DC machine system.

The use of the proposed method for simultaneous estimation combines many advantages. We don't need to use the speed and temperature sensors, the armature temperature estimation may be used for thermal condition monitoring, and the estimate of speed can be used on speed drive process. The resistance estimation may be used in adaptive calculations in the goal to escape the maladjustment phenomenon of the control by parameter variations such as the PID gain correction. The proposed estimator is suitable both in the drive and in the thermal monitoring.

In section 2, a thermal model of DC motor is presented. In section 3, the DC motor model has been resolved and some simulation results have been presented. In section 4, the ANN topology and design steps have been introduced. In section 5, the simulation studies of ANN estimator is carried out to verify and validate the convergence, effectiveness and estimation quality.

## II. Thermal Model of DC Motor

The model used in this paper is illustrated in [1].

### A. Electrical equation

$$V_a = R_{a0}(1+\alpha\,\theta)i_a + l_a \frac{di_a}{dt} + k_e\omega \tag{1}$$

Where *Va* is armature voltage, $R_{a0}$ is armature resistance at ambient temperature, $\alpha$ temperature coefficient of resistance, $\theta$ temperature above ambient, $i_a$ armature current, $l_a$ is armature inductance, *ke* is torque constant, and ω armature speed.

### B. Mechanical equation

$$T = k_e i_a = b\omega + J\frac{d\omega}{dt} + T_l \tag{2}$$

Where *b* is viscous friction constant, *J* is total inertia and $T_l$ is load torque.

### C. Thermal equations

The thermal model is derived by considering the power dissipation and heat transfer [25]. The power dissipated by the armature current flowing through the armature resistance, which varies in proportion to the temperature can be represented by:

$$P_j = R_{a0}(1+\alpha\,\theta)i_a^2 \tag{3}$$

The iron loss is proportional to speed squared for constant excitation, this loss variation with speed in the armature body can represent by:

The iron loss is proportional to speed squared for constant excitation multiplied by the iron loss constant $k_{ir}$ , this loss variation with speed in the armature body can represent by:

$$P_{ir} = k_{ir}\omega^2 \tag{4}$$

The power losses include contributions from copper losses and iron losses which frequency dependent:

$$P_l = R_{a0}(1+\alpha\,\theta)i_a^2 + k_{ir}\omega^2 \qquad (5)$$

A simple representation of the assumed DC machine heat flow is given in Fig. 1. Heat flow from the DC motor is either directly to the cooling air with heat transfer coefficient *k*.

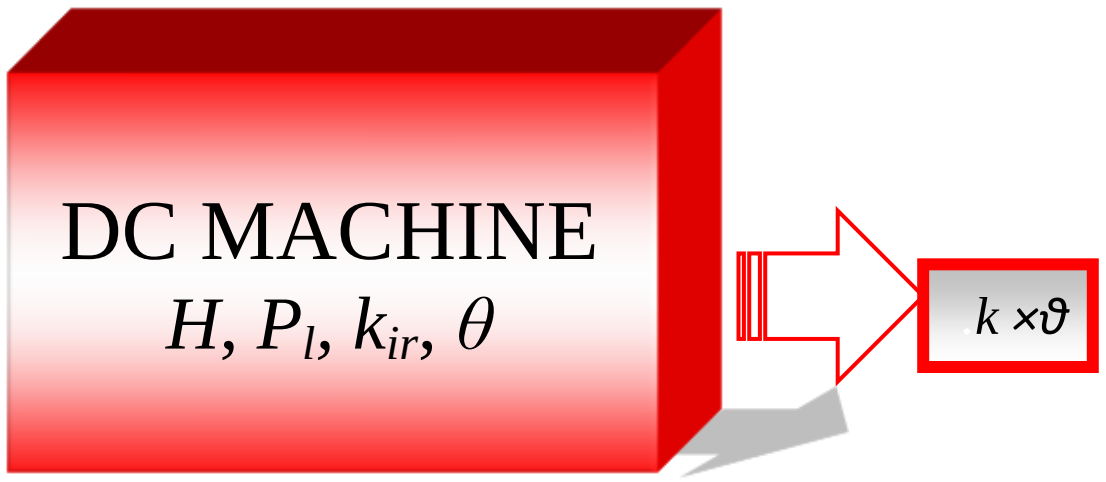

Fig. 1. Structure of thermal model of DC motor

The thermal power flow from the DC motor surface that is proportional to the difference temperature between the motor and the ambient air temperature, and the temperature variation in the armature which depends on the thermal capacity *H*.

$$P_l = k\,\theta + H\frac{d\theta}{dt} \qquad (6)$$

The effect of the cooling fan is approximated by introducing a speed dependence of the thermal transfer coefficient $k_T$ .

$$k = k_0(1+k_T\,\omega) \qquad (7)$$

When $K_o$: thermal transfer coefficient at zero speed and is $K_T$ thermal transfer coefficient with speed.

By arranging the previous eqs, we can write:

$$R_{a0}(1+\alpha\theta)i_a^2 + k_{ir}\omega^2 = k_0(1+k_T\,\omega)\theta + H\frac{d\theta}{dt} \qquad (8)$$

The equations system can be written as:

$$\frac{di_a}{dt} = -\frac{R_{a0}(1+\alpha\theta)}{l_a}i_a - \frac{k_e}{l_a}\omega + \frac{1}{l_a}V_a$$

$$\frac{d\omega}{dt} = \frac{k_e}{J}i_a - \frac{b}{J}\omega - \frac{1}{J}T_l$$

$$\frac{d\theta}{dt} = \frac{R_{a0}(1+\alpha\theta)}{H}i_a^2 + \frac{k_{ir}}{H}\omega^2 - \frac{k_0(1+k_T\omega)}{H}\theta \qquad (9)$$

## III. Simulation Results

The resolution of the equations system (9) in Matlab/Simulink environment with the use of parameters from [1], we get the following results:

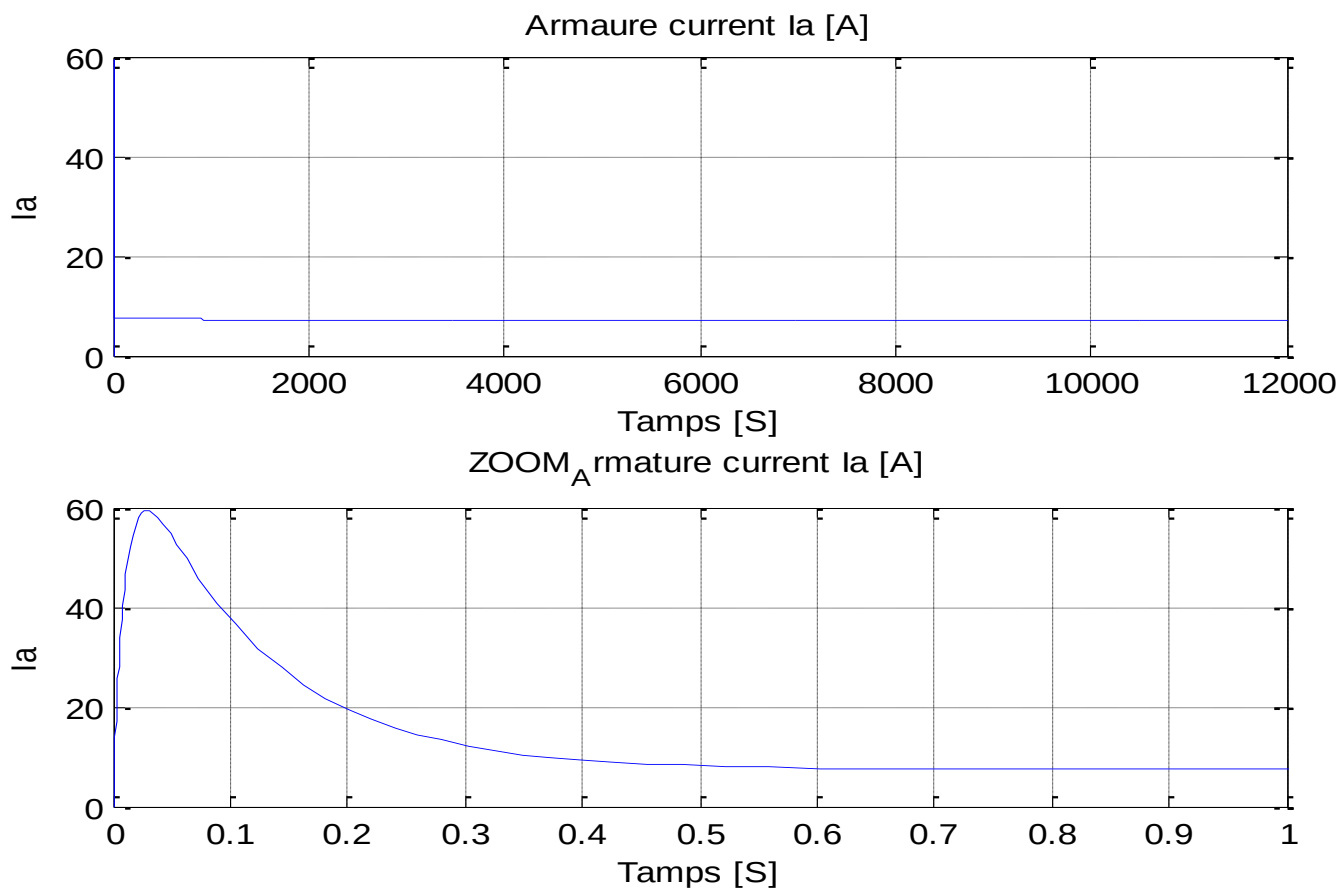

Fig. 2. Armature current

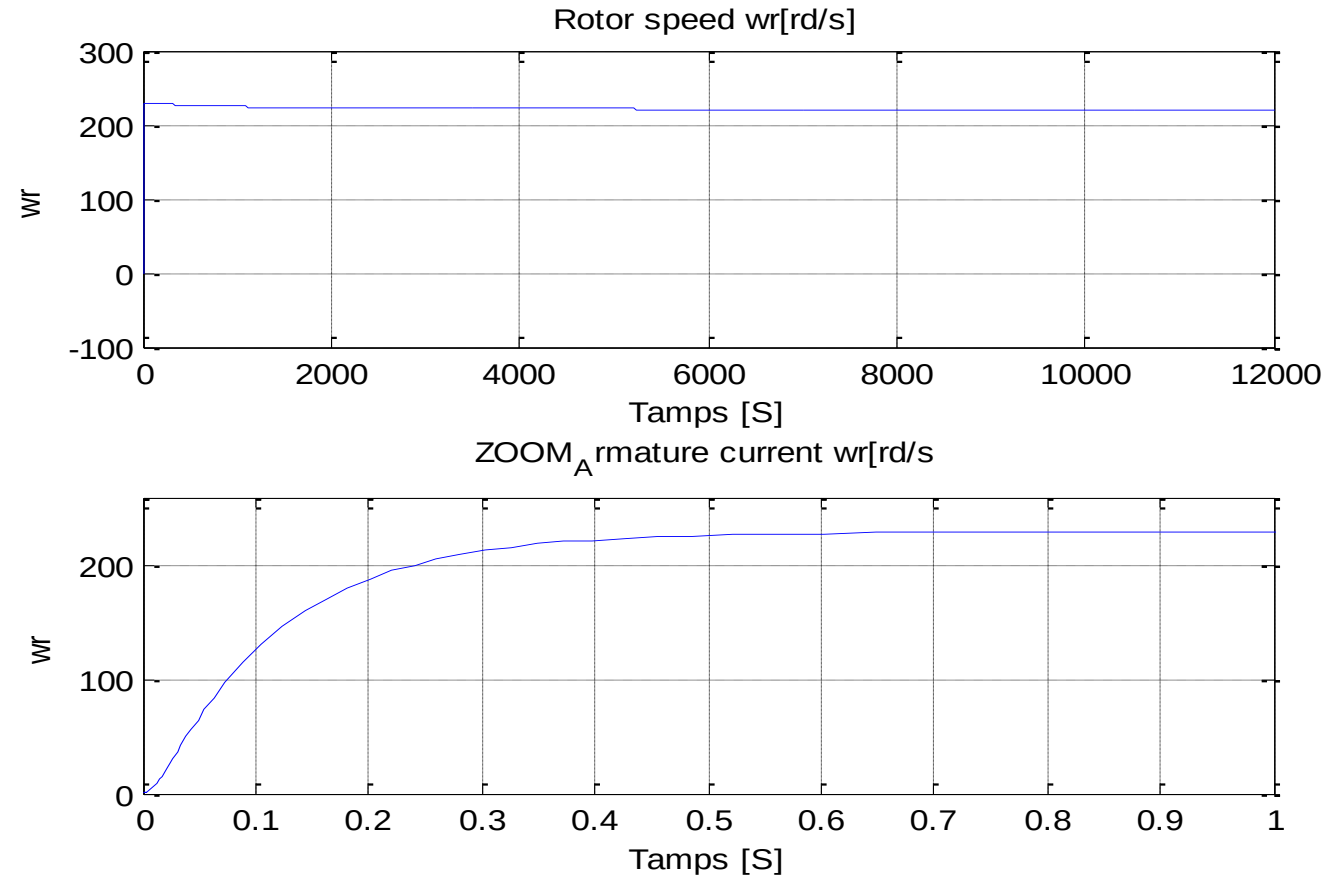

Fig. 3. Rotor speed

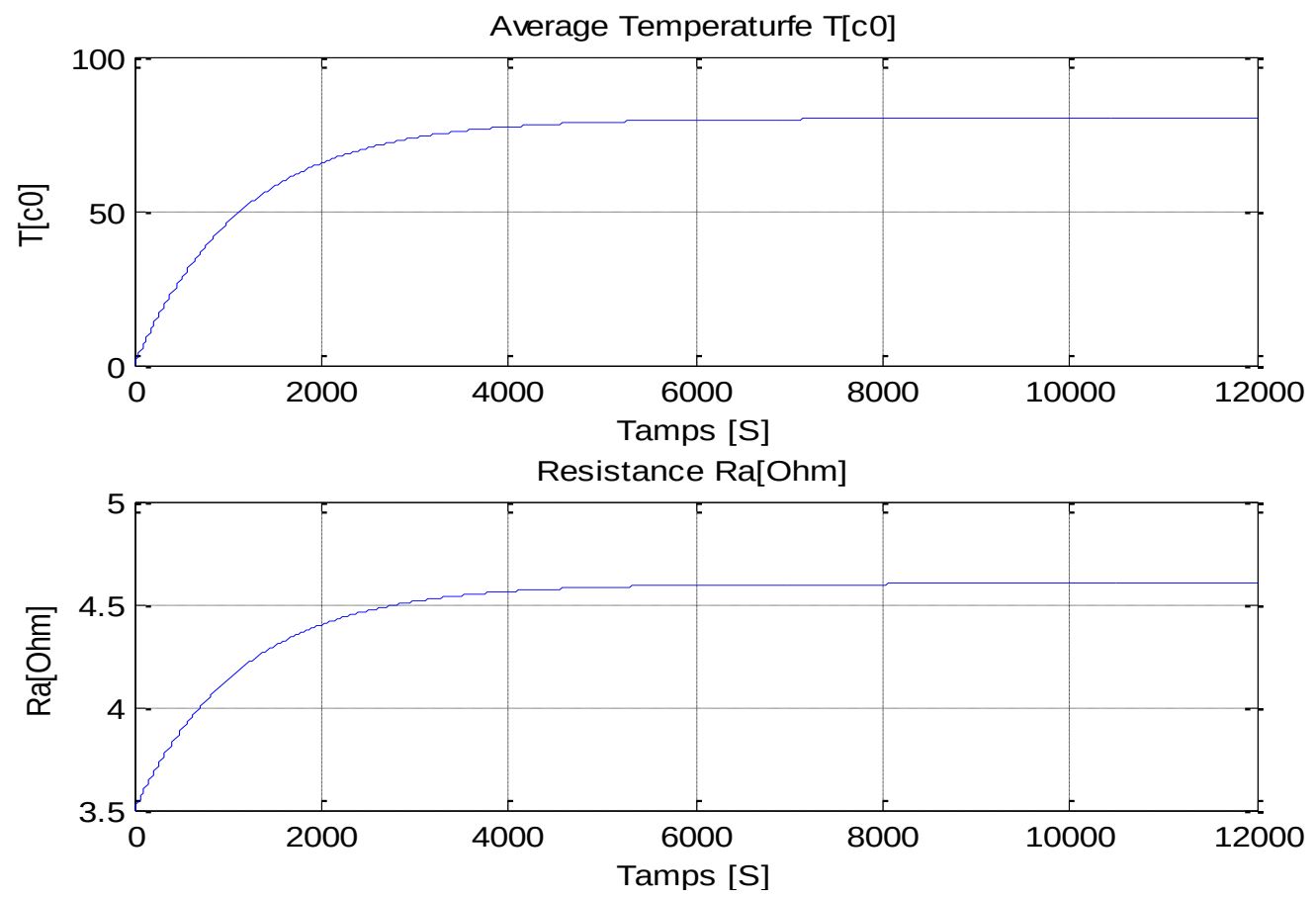

Fig. 4. Average temperature

The current curve variation is illustrated by Fig.2, we can see that in the transient stat the current reach 60A, but in the study state decrease by almost a factor of 10, the final value is 7.27A.

Fig.3 shows DC machine speed variation under load. Fig.4 shows armature average temperature in a brushed DC machine,

this temperature reaches 80 $^{0}$C after 140 min, the armature resistance increase 31%.

## IV. ANN ESTIMATOR

In this section, an ANN is used in tree steps in order to estimate the speed, temperature and resistance. In this section, we discuss the ANN design step, topology choice and the learning algorithms finally, we applicate the ANN to our study.

### A. *Types of ANN*

Feed-Forward Neural Network (FFNN) is the simplest process neural network. Each subsequent layer in FFNN only has a weight coming from the previous layer. Due to the drawback of this topology structure, FFNN cannot solve some complex problems [38]. The convergence process is slow or even impossible to realize. To address these problems, a CFNN is proposed here.

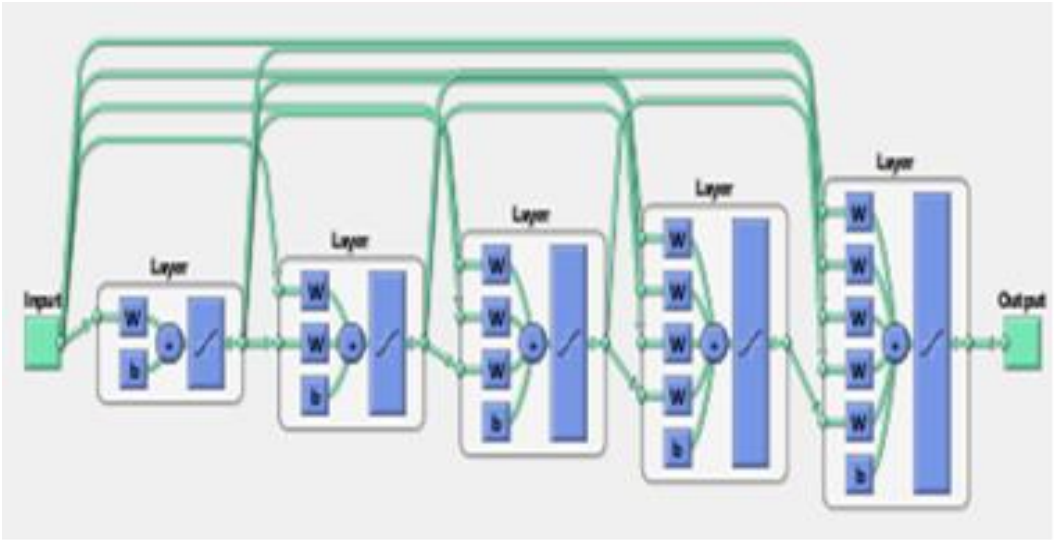

Fig. 5. The structure of the ANN used

CFNN are similar to FFNN, but include a connection from the input and every previous layer to following layers. As with FFNN, a two-or more layer cascade-network can learn any finite input-output relationship arbitrarily well given enough hidden neurons [38-29].

### B. *Data sets*

We have create a Matlab program that breaks the input vector into three parts without losing the information of each part, to make the data obtained by simulation similar than the sensor data a random white-Gaussian noise signal has been added.

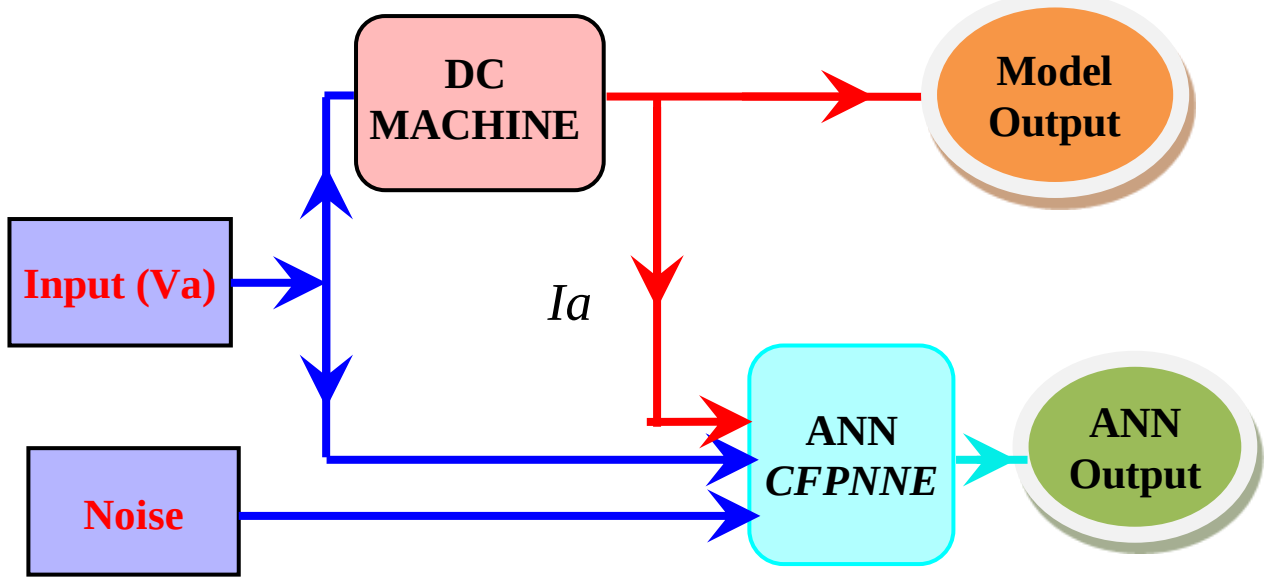

Fig. 6. ANN estimator schemes

This noise make the training very hard and requires a significant time, but the ANN is very trained and applicable on real time. so we have three sets: training, test and validation, each base part in the input vector of a well-defined percentage, 50% occupied by training set, 25% by the testing and 25% by validation set, this data was extracted from Fig. 6.

### C. *Training*

LMBP is the default training function because it is very fast, but it requires a lot of memory to run [38-40]. In our case, we have a very large input vector so the problem of exceed memory is imposed.

We have created a Matlab program for optimize CFNNE performances, such as hidden layer number, number of neurons in each hidden layer, epochs number. For the activation functions, we try deferent functions, but the hyperbolic tangent sigmoid transfer function for the hidden layers and linear transfer function for the output are the best.

Fig. 7 shows ANN estimator used in the present paper.

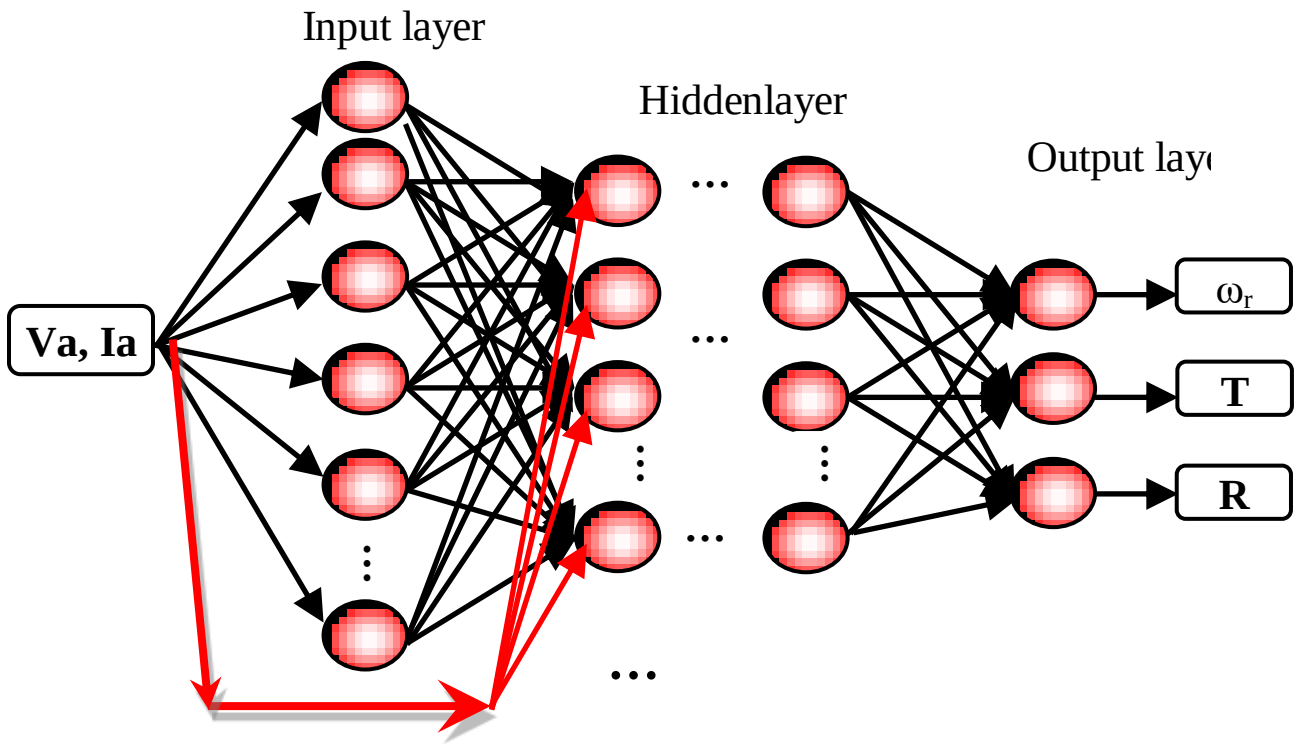

Fig. 7. ANN estimator used in the present paper

## V. SIMULATION RESULTS

In this section we following the instructions discussed in the past section for obtained an optimized CFNNE, training step is the most important step to create any ANN, our optimized CFNNE is trained after 2000 epoch at the performance 1.6e-4.

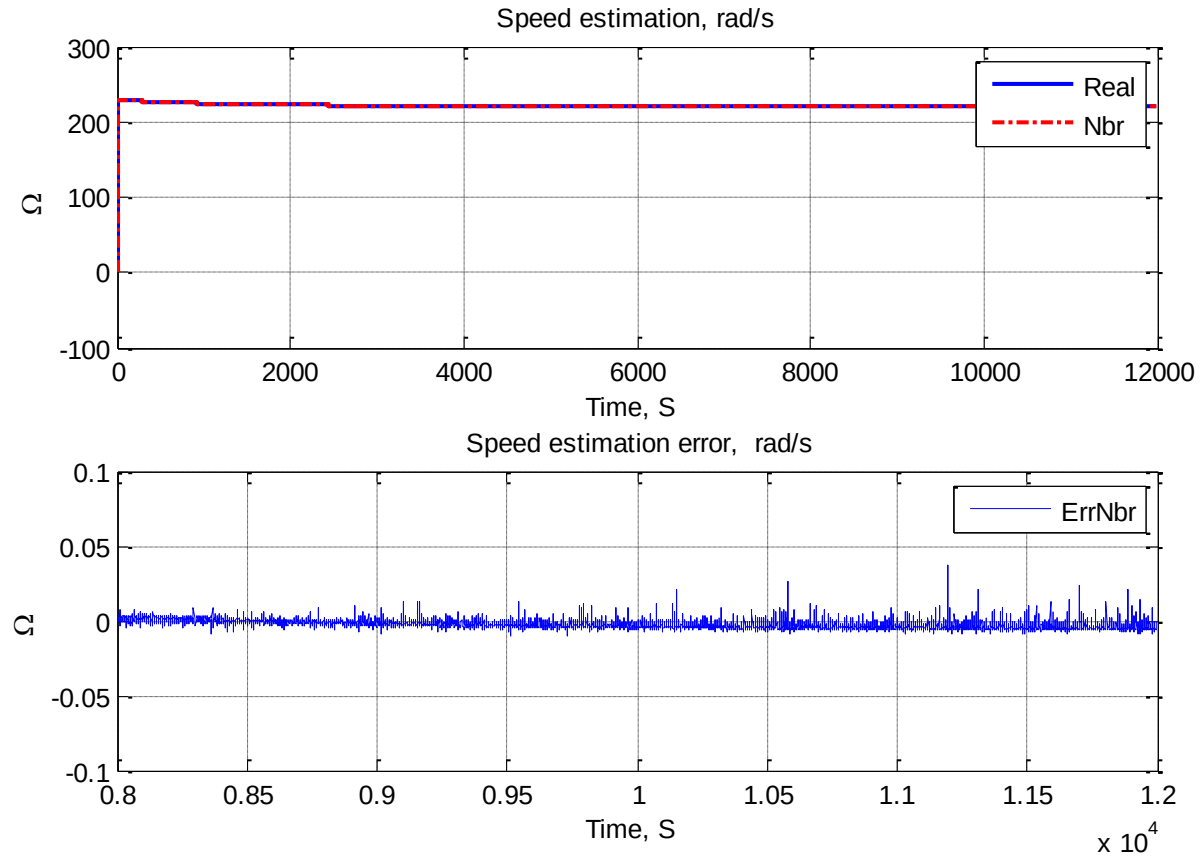

Fig. 8. Speed estimation by ANN

Fig.8 shows DC machine speed estimation and the corresponding estimation error at the testing step, in transient state we can see on the speed estimation error curve's a peak of 110 rad/s between the output of the model and the ANN output, the duration of this peak is 0.3s. In steady state our CFNNE give a good results with estimation error less than 0.04 rad/s that means less than 0.008%.

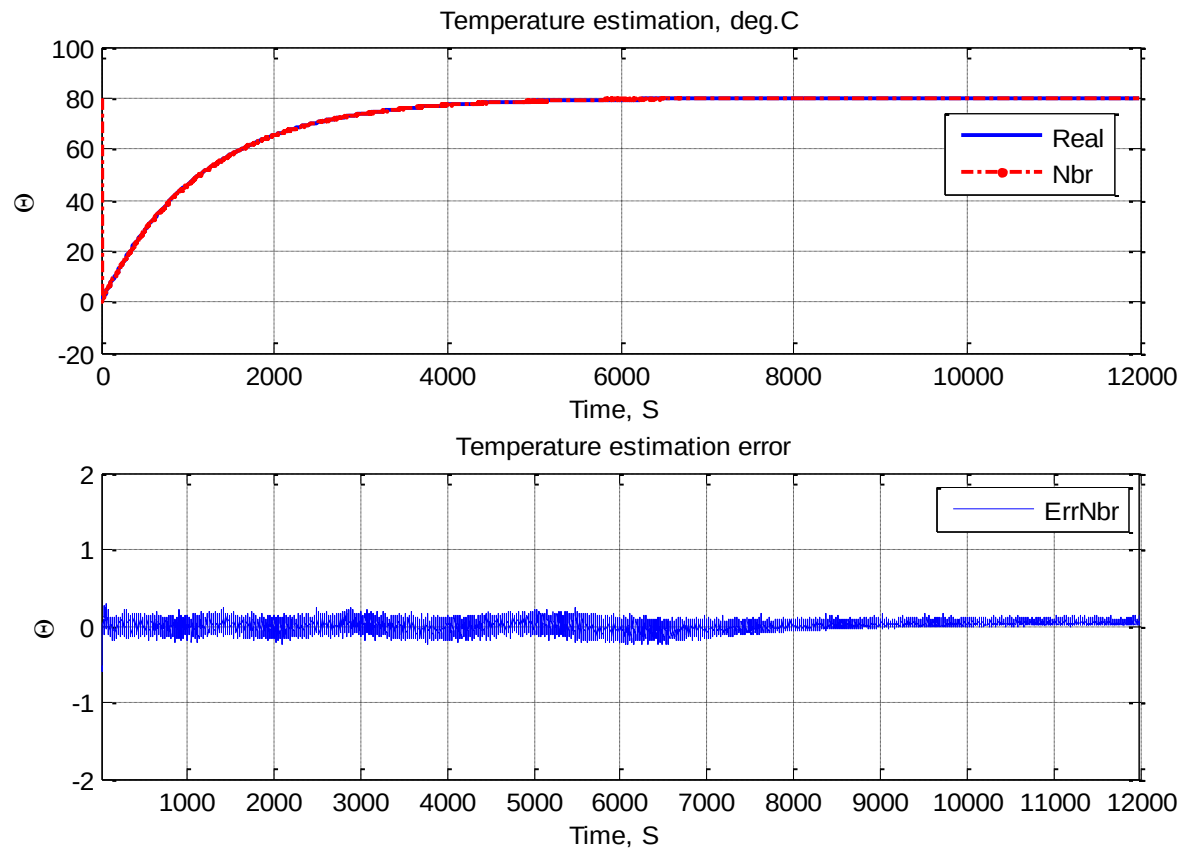

Fig. 9. Average temperature estimation by ANN

The temperature estimation is shown in Fig.9; the temperature value in the DC machine thermal steady state is approximately $80^0$C, the corresponding error is less than $0.6^0$C so, less than 0.75%.

The CFNNE can also estimate the resistance, this estimation is shown in Fig. 10 the estimation error is less than 0.004Ω.

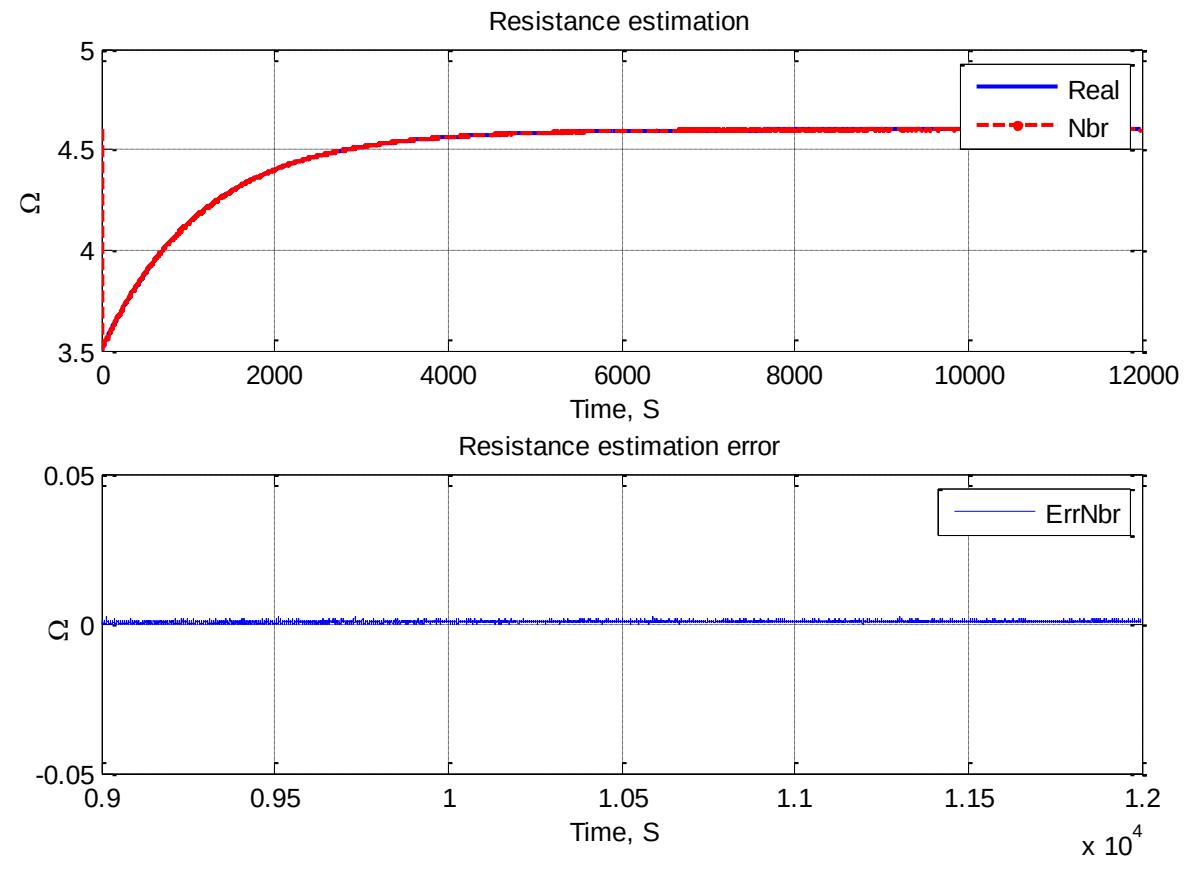

Fig. 10. Armature resistance estimation by ANN

## VI. Conclusion

A thermal model of DC motor is presented and some results are discussed. The measurement problems and even the use of conventional estimators of speed, temperature and resistance were discussed; the simultaneous estimation of DC machine state variables and parameters not recognized in the literature, our goal is to simulate simultaneously the DC machine speed, temperature and resistance. The ANN makes it possible to achieve this goal, because it enables to estimate simultaneously the speed, temperature and resistance of a DC motor from only the knowledge of voltage and current. The specialized literature we give several ANN versions, according to the studied system characteristics the most suitable approach is CFNNE. The creation steps of CFNNE and the different data bases is discussed in Section IV, the addition of white Gaussian noise to the data set is very important, because if that which make the application in real time is possible and our ANN not be affected by current and voltage measurements noise. It can be seen that the network has worked with an acceptable error. The variable stat estimation may be used in condition monitoring or in robust control, the simulation results demonstrate that the new approach proposed in this paper is feasible.

## References

[1] P. P. Acarnley, J. K. Al-Tayie, Estimation of speed and armature temperature in a brushed DC drive using the extended Kalman filter, IEE Proc Electr. Power Appl., vol. 144, no. 1, pp. 13–20, Jan I997.

[2] E. Fiorucci, G. Bucci, F.Ciancetta, D. Gallo, C. Landi and M. Luiso, variable speed drive characterization: review of measurement techniques and future trends, Advances in Power Electronics, vol. 2013, pp.1–14, 2013.

[3] G. Bucci, C. Landi, Metrological characterization of a contactless smart thrust and speed sensor for linear induction motor testing, Instrumentation and Measurement, IEEE Transactions on , vol. 45, no.2, pp. 493 – 498, Apr 1996.

[4] R. J. Welch and G. W. Younkin, How Temperature Affects a Servomotor's Electrical and Mechanical Time Constants, Proc. IEEE Ind. Appl. Conference, vol. 2, pp. 1041–1046, 13-18 Oct. 2002.

[5] IEEE Recommended Practice for General Principles of Temperature Measurement as Applied to Electrical Apparatus, IEEE Std 119-1974,1974.

[6] T. Chunder, Temperature rise measurement in armature of a DC motor, under running conditions by telemetry, Proc. Sixth International Conference on Electrical Machines and Drives, pp. 44–48, 8-10 Sep 1993.

[7] L. Michalski, K. Eckersdorf, J. Kucharski, J. McGhee, Temperature Measurement, John Wiley & Sons Ltd, 2001.

[8] I. J. Aucamp, L .J. Grobler, Heating, ventilation and air conditioning management by means of indoor temperature measurements, Proc. 9th conference industrial and commercial use of energy (ICUE), pp. 1–4, 15-16 Aug, 2012.

[9] A. Cassat, C. Espanet and N. Wavre, BLDC Motor Stator and Rotor Iron Losses and Thermal Behavior Based on Lumped Schemes and 3-D FEM Analysis, IEEE Transactions on Industry Applications, vol. 39, no. 5, pp. 1314–1322, 2003.

[10] J. Le Besnerais, A. Fasquelle, M. Hecquet, J. Pellé, V. Lanfranchi, S. Harmand, P. Brochet and A. Randria, Multiphysics Modeling: Electro-Vibro-Acoustics and Heat Transfer of PWM-Fed Induction Machines, IEEE Transactions on Industrial Electronics, vol. 57, no. 4, pp. 1279–1287, 2010.

[11] R. Lazarevic, P. Radosavljevic, A. Osmokrovic, novel approach for temperature estimation in squirrel-cage induction motor without sensors, IEEE Transactions on Instrumentation and Measurement, vol. 48, no. 3, pp. 753–757, 1999.

[12] S. B. Lee, T. G. Habetler, R. G. Harley and D. J. Gritter, A stator and rotor resistance estimation technique for conductor temperature monitoring, Proc. IEEE Ind. Appl. Conference, vol. 1, pp. 381–387, 2000.

[13] S. B. Lee, T. G. Habetler, R. G. Harley and D. J. Gritter, An Evaluation of Model-Based Stator Resistance Estimation for Induction Motor Stator Winding Temperature Monitoring, IEEE Transactions on Energy Conversion, vol. 17, no. 1, pp. 7–15, 2002.

[14] S. B. Lee, T. G. Habetler, An Online Stator Winding Resistance Estimation Technique for Temperature Monitoring of Line-Connected Induction Machines, IEEE Transactions on Industry Applications, vol. 39, no. 3, pp. 685–694, 2003.

[15] K. D. Hurst, T.G. Habetler, A thermal monitoring and parameter tuning scheme for induction machines, Proc. IEEE Ind. Appl. Conference, IEEE-IAS Annu. Meeting, vol. 1, pp. 136–142, 1997.

[16] H. Mellah, K. E. Hemsas, Stochastic Estimation Methods for Induction Motor Transient Thermal Monitoring Under Non Linear Condition, Leonardo Journal of Sciences, vol. 11, pp. 95–108, 2012.

[17] J. F. Moreno, F. P. Hidalgo and M. D. Martinez, Realisation of tests to determine the parameters of the thermal model of an induction machine, IEE Proc Electr. Power Appl., vol. 148, no.5, pp. 393–397, 2001.

[18] R. Beguenane, M.E.H. Benbouzid, Induction motors thermal monitoring by means of rotor resistance identification, IEEE Transaction on Energy Conversion, vol. 14, no. 3, pp. 566-570, 1999.

[19] M.S.N. Saïd, M.E.H. Benbouzid, H–G Diagram Based Rotor Parameters Identification for Induction Motors Thermal Monitoring, IEEE Transactions on Energy Conversion, vol. 15, no. 1, pp. 14–18, 2000.

[20] Z. Gao, T. G. Habetler, R. G. Harley and R. S. Colby, An Adaptive Kalman Filtering Approach to Induction Machine Stator Winding Temperature Estimation Based on a Hybrid Thermal Model, Proc. IEEE Ind. Appl. Conference, IEEE-IAS Annu. Meeting, vol. 1, pp. 2–9, 2005.

[21] R. Pantonial, A. Kilantang and B. Buenaobra, Real time thermal estimation of a Brushed DC Motor by a steady-state Kalman filter algorithm in multi-rate sampling scheme, Proc TENCON 2012 IEEE Region 10 Conference, pp. 1–6, 19-22 Nov 2012.

[22] W. Zhang, S. G. Andrew and R.H. Saeid, Nonlinear Estimation of Stator Winding Resistance in a Brushless DC Motor, Proc American Control Conference (ACC), pp. 4699-4704, 17-19 June 2013.

[23] M. Jabri, I. Chouire and N.B. Braiek, Fuzzy Logic Parameter Estimation of an Electrical System, Proc. International Multi-Conference on Systems, Signals and Devices, pp.1–6, 2008.

[24] M. Jabri, A. Belgacem and Houssem Jerbi, Moving Horizon Parameter Estimation of Series Dc Motor Using Genetic Algorithm, Proc. International Multi-Conference on Systems, Signals and Devices, pp. 26–27, 2009.

[25] S. A. Kalogirou, Artificial neural networks in renewable energy systems applications: a review, Renewable and Sustainable Energy Reviews, vol. 5, no. 4, pp.373–401, 2001.

[26] E. Byvatov, U. Fechner, J. Sadowski and G. Schneider, Comparison of support vector machine and artificial neural network systems for drug/nondrug classification, Journal of Chemical and modeling, vol. 43, no. 6, pp. 1882–1889, 27 Sept, 2003

[27] S. Agatonovic-Kustrin, R. Beresford, Basic concepts of artificial neural network (ANN) modeling and its application in pharmaceutical research, Journal of pharmaceutical and Biomedical Analysis, vol. 22, no. 5, pp. 717–727, 2000.

[28] S. Ablameyko, L.Goras, M. Gorz and V. Piuri, Neural Networks for Instrumentation, Measurement and Related Industrial Applications, IOS Press, 2003.

[29] S. Haykin, Kalman filtering and neural networks, John Wiley & Sons, 2001.

[30] A. Cochocki, R. Unbehauen, Neural networks for optimization and signal processing. John Wiley & Sons, Inc, 1993.

[31] M. Y. Chow, Y. Tipsuwan, Neural plug-in motor coil thermal modeling, in Industrial Electronics Society, 2000. IECON 2000. 26th Annual Conference of the IEEE, vol.3, no., pp.1586–1591, 2000.

[32] L. P. Veelenturf, Analysis and applications of artificial neural networks, Prentice-Hall, Inc., 1995.

[33] M. Gupta, L. Jin and N. Homma, Static and dynamic neural networks: from fundamentals to advanced theory, John Wiley & Sons, 2004.

[34] L. C. Jain, N.M. Martin, Fusion of Neural Networks, Fuzzy Systems and Genetic Algorithms: Industrial Applications, vol. 4, CRC press. 1998.

[35] R. C. Eberhart, J. Kennedy, A New Optimizer Using Particle Swarm Theory, Proceedings of the Sixth International Symposium on Micro Machine and Human Science, MHS '95, vol.1, pp. 39–43. 1995.

[36] J.S.R. Jang, C.T. Sun and E. Mizutani, Neuro-Fuzzy and Soft Computing: A Computational Approach to Learning and Machine Intelligence, Prentice-Hall, 1997.

[37] C. Dimoulas, G. Kalliris, G. Papanikolaou, V. Petridis and A. Kalampakas, Bowel-sound pattern analysis using wavelets and neural networks with application to long-term, unsupervised, Expert Systems with Applications, vol. 34, no. 1, pp. 26–41, 2008.

[38] B.M. Wilamowski, How to not get frustrated with neural networks, Proc. IEEE Int. Conf. Ind. Technol, pp. 5–11., 2011.

[39] Zhou Yao-ming, Meng Zhi-jun, Chen Xu-zhi and Wu Zhe, Helicopter Engine Performance Prediction based on Cascade-Forward Process Neural Network, IEEE Conference on Prognostics and Health Management (PHM), pp, 1–5, 18-21 June 2012.

[40] H. Demuth, M. Beale and M. Hagan, Neural Network Toolbox Users Guide, the MathWorks, Natrick, USA. 2009.